\documentclass[sigconf,nonacm,margin=0.8in]{acmart}

\usepackage{lipsum}
\usepackage{tabularx}
\usepackage{booktabs}
\usepackage{array}
\usepackage[table,xcdraw]{xcolor}
\usepackage{pgfplots}
\usepackage{xcolor}
\usepackage{dblfloatfix}
\usepackage{balance} 

\pgfplotsset{compat=1.18} 

\begin{document}
\title{Can You Trust Your Copilot? A Privacy Scorecard for AI Coding Assistants}

\author{Amir AL-Maamari}
\email{almaam03@ads.uni-passau.de}
\affiliation{
  \institution{University of Passau}
  \country{Germany}
}

\begin{abstract}
The rapid integration of AI-powered coding assistants into developer workflows has raised significant privacy and trust concerns. As developers entrust proprietary code to services like OpenAI's GPT, Google's Gemini, and GitHub Copilot, the unclear data handling practices of these tools create security and compliance risks. This paper addresses this challenge by introducing and applying a novel, expert-validated privacy scorecard. The methodology involves a detailed analysis of four document types—from legal policies to external audits—to score five leading assistants against 14 weighted criteria. A legal expert and a data protection officer refined these criteria and their weighting. The results reveal a distinct hierarchy of privacy protections, with a 20-point gap between the highest- and lowest-ranked tools. The analysis uncovers common industry weaknesses, including the pervasive use of opt-out consent for model training and a near-universal failure to filter secrets from user prompts proactively. The resulting scorecard provides actionable guidance for developers and organizations, enabling evidence-based tool selection. This work establishes a new benchmark for transparency and advocates for a shift towards more user-centric privacy standards in the AI industry.
\end{abstract}


\maketitle
\thispagestyle{plain}
\pagestyle{plain}

\section{Introduction}
\label{sec:introduction}

AI-powered coding assistants like GitHub Copilot and Amazon Q Developer have rapidly integrated into developer workflows, becoming essential programming tools \cite{generativeAIAssistants2025}. A recent survey highlights their utility: 53\% of developers use them for boilerplate code, and 43\% for debugging \cite{codesignal2024}. This reliance on remote processing, however, creates privacy risks, as proprietary code and sensitive data are transmitted to third-party servers. High-profile incidents, such as Samsung banning ChatGPT after confidential code leaks \cite{petkauskas2023samsung, degeurin2023samsung} and Italy's GDPR-related suspension of the service \cite{ferrari2023garante}, highlight the conflict between utility and privacy. While developers embrace these tools for productivity, the lack of transparent privacy practices creates a trust deficit.

\subsection{Problem Statement}
The central problem is the opacity of privacy practices across competing AI coding assistants. Providers publish lengthy and jargon-laden legal documents, preventing developers and organizations from answering fundamental questions: Which tool can be trusted with confidential code? What happens to submitted data? Without clear, systematic comparisons, organizations risk adopting tools that violate compliance requirements or internal intellectual property (IP) policies by storing code indefinitely or using it for model retraining. While existing research has explored broader LLM ethics and security \cite{madampe2025aibasedprogrammingassistantsprivacyrelated,wang2025trustworthy}, no study provides a standardized, comparative privacy analysis for the coding assistants developers use daily. This work addresses this gap by providing a standardized framework and analysis.

\subsection{Research Goal}

This paper evaluates the privacy practices of five prominent AI coding assistants—OpenAI's GPT models, Anthropic Claude, Google Gemini, GitHub Copilot, and Amazon Q Developer—using a comparative privacy scorecard. The analysis is grounded in a comprehensive evidence corpus drawn from each provider's core legal documents, enterprise-tier agreements, technical documentation, and external audits. The study is guided by the following research questions:
\begin{enumerate}
    \item RQ1: What data do popular AI coding assistants collect from users’ coding sessions, and how is this information stored or retained?
    \item RQ2: In what ways do these coding assistants use or share user-provided code (e.g., for model training or with third parties), and what privacy controls (opt-outs, data deletion, etc.) are offered to users?
    \item RQ3: How do the privacy provisions of the coding assistants compare when evaluated against a common set of criteria? Which assistants emerge as more privacy-preserving, and which have notable weaknesses?
    \item RQ4: What are the implications of these comparative findings for software practitioners and policymakers? For example, how should organizations select AI coding tools under privacy constraints, and what improvements are needed in industry practices?
\end{enumerate}

To answer these questions, this study developed and applied an expert-validated privacy scorecard that evaluates each assistant against different criteria, including data governance, technical safeguards like secret filtering, and the transparency of user controls. The evaluation is strictly document-focused, assessing providers' stated policies without performing penetration testing or code auditing.

\subsection{Contributions}

This work makes the following contributions to the field of privacy in AI-assisted software engineering:
\begin{enumerate}
    \item \textbf{An Expert-Validated Privacy Scorecard Framework:} Introduction of a novel, reusable framework with specific, weighted criteria to assess the privacy posture of AI coding assistants.

    \item \textbf{A Comparative Analysis of Leading Tools:} Application of the scorecard to five market-leading assistants, revealing key differences in their data handling practices and surfacing undocumented privacy risks.

    \item \textbf{Actionable Rankings and Guidance:} A clear privacy ranking of the evaluated tools and concrete recommendations for developers, organizations, and policymakers to mitigate risks and foster trustworthy AI development practices.
\end{enumerate}

\section{Related Work}
\label{sec:related-work}

The rapid adoption of LLM-based coding tools has sparked extensive discussion about privacy, security, and intellectual property implications \cite{xu2024first}. Several studies have begun examining how well the data practices of these AI services align with legal requirements. For instance, \cite{vu2025comparative} conducted a structured analysis of major LLM providers’ privacy policies (covering models like OpenAI’s ChatGPT, Google’s Bard/Gemini, Microsoft’s Copilot, Anthropic’s Claude, etc.) against data protection laws such as the GDPR. Their work produced privacy compliance scores for each model and identified potential legal compliance risks in current policy statements. Similarly, \cite{chadwick2024assessments} rigorously evaluated the privacy compliance of popular commercial LLMs (focusing on ChatGPT and Claude) and found gaps between the statements of the policies and the actual data protection measures. These legal and policy analyses provide valuable high-level benchmarks on LLM privacy; however, they focus on general-purpose models and broad regulations, rather than the niche context of coding assistants. 

Another line of relevant research looks at the trust and usage of AI assistants among software developers. \cite{madampe2025doingusingaibasedprogramming} surveyed practitioners about using AI-based programming assistants for privacy-related coding tasks, finding that more than 64\% of the respondents distrusted the ability of these tools to handle privacy requirements. The primary reasons cited were lack of transparency and unpredictable behavior, leading many developers to double-check or avoid AI suggestions when sensitive data was involved. This aligns with broader findings that emphasize transparency and reliability as key to trusting AI code completions. In addition, the security of code generated by assistants has come under scrutiny. Previous work by \cite{10.1145/3610721} showed that GitHub Copilot can produce vulnerable code in about 40\% of cases under certain conditions, highlighting that trust in such tools is multifaceted - covering not just privacy of input data, but also quality of output. There have even been studies on the memorization risks of LLMs, demonstrating that models can regurgitate training data verbatim (including sensitive snippets of code or personal information) if prompted cleverly as \cite{274574} evaluated. Such results underscore why unbridled data collection by coding assistants (for example, using user code to retrain models) is worrisome from a privacy standpoint. 

This paper builds on these previous efforts, but diverges in important ways to fill the identified gap. First, the work focuses specifically on AI coding assistants, a domain with unique risks. When the data in question are proprietary source code, privacy lapses translate directly into intellectual property leaks and security vulnerabilities, concerns not fully addressed by analyses of general chatbot policies. Second, the methodology goes beyond providers’ high-level privacy policies by mining a broader evidence corpus of documents (technical FAQs, enterprise agreements, compliance certifications, and so on) to verify how privacy practices are operationalized. This yields a more holistic assessment than legal compliance checklists alone. Finally, whereas prior studies stop at qualitative findings or compliance scores, it contributes an expert-validated quantitative scorecard tailored to coding assistants. The scorecard’s weighted criteria translate complex policy details into an actionable ranking. In summary, it shifts from general analysis of LLM privacy toward a specialized, practitioner-focused comparison.

\section{Methodology}
This research follows a three-phase methodology to develop and apply a privacy scorecard for AI coding assistants. First, a comprehensive evidence corpus was collected. Second, the scorecard framework was developed and validated with industry experts. Third, the selected assistants were quantitatively scored and ranked.

\subsection{Phase 1: Evidence Corpus Collection}
The analysis is grounded in a comprehensive evidence corpus collected for each assistant and snapshotted from mid-2025 to ensure comparability. This corpus spans four distinct, predefined categories of documentation to provide a holistic view of each provider's privacy posture:
\begin{enumerate}
    \item \textbf{Core Legal Documents}: Public-facing policies governing general use, such as the Privacy Policy and Terms of Service. These establish the baseline legal relationship with the user.
    \item \textbf{Enterprise-Tier Documents}: Commitments made to paying business customers, found in documents like Data Processing Addendums (DPAs), Enterprise Agreements, and Trust Center pages. These often contain stronger privacy guarantees.
    \item \textbf{Technical and Developer-Facing Documentation}: Materials explaining the service's technical operation, including developer guides, API documentation, white papers, and model cards. These reveal practical implementation details.
    \item \textbf{External and Verifiable Evidence}: Independent, third-party sources that verify or challenge provider claims, such as SOC 2 audit reports, ISO certifications, and public records of regulatory fines or data breaches.
\end{enumerate}
For each evaluation criterion, the methodology predefined the primary document sources to consult and specific keywords to locate relevant evidence, ensuring every score is directly traceable to a specific, documented passage.

\subsection{Phase 2: Framework Development and Validation}
Grounded in the evidence corpus and established privacy principles (such as those in GDPR, e.g., purpose limitation and data minimization), this study developed an evaluation framework comprising 14 sub-criteria organized into three categories. The initial criteria were derived from these principles and then refined based on the specific features and risks identified within the collected documents.

The validation process involved a legal technical expert and a technical data protection officer (DPO). Using a structured survey, they assessed each sub-criterion for \textit{Relevance} and \textit{Clarity} on a 5-point scale. To determine the category weights, the experts were asked to perform a \textbf{100-point constant-sum allocation}, distributing points across the three main categories based on their professional judgment of which area posed the greatest risk to user privacy.

This expert-driven process produced the final, validated scorecard framework, presented in \textbf{Table~\ref{tab:final-framework}}. The weights assigned by experts confirm that \textbf{Technical Safeguards (42.5\%)} are considered the most critical risk area, followed by \textbf{Data Governance \& Legal Compliance (32.5\%)} and \textbf{Transparency \& User Control (25.0\%)}. The experts rated all criteria as highly relevant (average 5.0/5.0) and clear (average 4.8/5.0), confirming the framework's robustness.

\begin{table*}[t]
\centering
\caption{Validated Framework Table showing average expert scores across criteria, with expert comments provided only at the category level.}
\label{tab:final-framework}
\renewcommand{\arraystretch}{1.20} 
\begin{tabularx}{\textwidth}{>{\raggedright\arraybackslash}p{0.32\textwidth} >{\centering\arraybackslash}p{0.08\textwidth} >{\centering\arraybackslash}p{0.08\textwidth} >{\centering\arraybackslash}p{0.1\textwidth} >{\raggedright\arraybackslash}X}
\toprule
\textbf{Criterion Category \& Sub-Criteria} & \textbf{Relevance} & \textbf{Clarity} & \textbf{Weight} & \textbf{Expert Comment} \\
\midrule

\textbf{A: Data Governance \& Legal Compliance} & \textbf{5.0} & \textbf{5.0} & \textbf{32.5\%} & \textbf{Confirmed as essential and clearly defined. Experts emphasized the importance of IP rights and explicit consent.} \\
A1: Public \& clear DPA (Data Processing Addendum) & 5.0 & 5.0 & \cellcolor{gray!20} & \cellcolor{gray!20} \\
A2: Explicit consent for training & 5.0 & 5.0 & \cellcolor{gray!20} & \cellcolor{gray!20} \\
A3: Clear data residency policy & 5.0 & 5.0 & \cellcolor{gray!20} & \cellcolor{gray!20} \\
A4: Definition of IP rights & 5.0 & 5.0 & \cellcolor{gray!20} & \cellcolor{gray!20} \\
A5: Straightforward DSAR (Data Subject Access Request) mechanisms & 5.0 & 5.0 & \cellcolor{gray!20} & \cellcolor{gray!20} \\
\midrule

\textbf{B: Technical Privacy \& Security Safeguards} & \textbf{5.0} & \textbf{4.6} & \textbf{42.5\%} & \textbf{Confirmed as the most critical category. Minor ambiguity in technical terms was noted, requiring precise definitions in the scoring rubric.} \\
B1: Anonymization for training data & 5.0 & 5.0 & \cellcolor{gray!20} & \cellcolor{gray!20} \\
B2: Role-Based Access Control (RBAC) & 5.0 & 4.0 & \cellcolor{gray!20} & \cellcolor{gray!20} \\
B3: Encryption in transit and at rest & 5.0 & 5.0 & \cellcolor{gray!20} & \cellcolor{gray!20} \\
B4: Filtering of secrets/PII & 5.0 & 5.0 & \cellcolor{gray!20} & \cellcolor{gray!20} \\
B5: Regular 3rd-party audits/pen-testing & 5.0 & 4.0 & \cellcolor{gray!20} & \cellcolor{gray!20} \\
\midrule

\textbf{C: Transparency \& User Control} & \textbf{5.0} & \textbf{5.0} & \textbf{25.0\%} & \textbf{Confirmed as essential and clearly defined. Experts valued granular user controls and clear data retention disclosures.} \\
C1: Granular opt-out controls & 5.0 & 5.0 & \cellcolor{gray!20} & \cellcolor{gray!20} \\
C2: Transparency on telemetry & 5.0 & 5.0 & \cellcolor{gray!20} & \cellcolor{gray!20} \\
C3: Clear data retention periods & 5.0 & 5.0 & \cellcolor{gray!20} & \cellcolor{gray!20} \\
C4: Publicly available model card & 5.0 & 5.0 & \cellcolor{gray!20} & \cellcolor{gray!20} \\
\bottomrule
\end{tabularx}
\end{table*}

\subsection{Phase 3: Quantitative Scoring and Ranking}

This study evaluated five leading proprietary AI coding assistants—\textbf{\allowbreak Amazon  Q Developer}, \textbf{Anthropic Claude}, \textbf{GitHub Copilot}, \textbf{Google Gemini}, and \textbf{OpenAI's GPT models}—selected for their market share and technological influence.

The scoring was conducted as follows:
\begin{itemize}
    \item \textbf{Evidence-Based Rubric.} Each assistant was scored against the 14 sub-criteria using a 3-point ordinal rubric (0 = Not Met, 1 = Partially Met, 2 = Fully Met).
    \item \textbf{Weighted Score Aggregation.} A final composite score for each tool, scaled to 100, was calculated using the expert-defined weights. The score is computed as:
    \[
    \text{Score}_{\text{tool}} = 100 \cdot \sum_{i \in \{A,B,C\}} (w_i \cdot s_i)
    \]
    where $w_i$ is the expert-defined weight for category $i$ and $s_i$ is the tool's normalized score for that category. A higher score indicates better adherence to the privacy best practices defined by the framework. The complete scoring rubric and evidence are available in a public replication package.
\end{itemize}

\section{Results and Discussion}

\definecolor{colorCatA}{HTML}{3182BD} 
\definecolor{colorCatB}{HTML}{31A354} 
\definecolor{colorCatC}{HTML}{FD8D3C} 

\begin{figure*}[t]
    \centering
    \pgfplotstableread[col sep=comma]{
    tool_id,display_name,catA,catB,catC,total_score
    google_gemini,Google Gemini,26.00,38.25,25.00,89.25
    anthropic_claude,Anthropic Claude,26.00,34.00,21.875,81.88
    github_copilot,GitHub Copilot,26.00,34.00,18.75,78.75
    amazon_q,Amazon Q,22.75,34.00,15.625,72.375
    openai_gpt,OpenAI GPT,19.50,29.75,18.75,68.0
    }\datatable
    \begin{tikzpicture}
        \begin{axis}[
            xbar stacked, width=\textwidth, height=5.2cm, bar width=11pt,
            xlabel={Final Privacy Score}, xmin=0, xmax=105, xtick={0,20,40,60,80,100},
            ytick=data,
            symbolic y coords={openai_gpt, amazon_q, github_copilot, anthropic_claude, google_gemini},
            yticklabels from table={\datatable}{display_name},
            y axis line style={opacity=0}, ytick style={draw=none}, enlarge y limits={0.18},
            legend style={at={(0.5, -0.2)}, anchor=north, legend columns=-1, draw=none, /tikz/every even column/.append style={column sep=0.5cm}},
            nodes near coords style={font=\small\bfseries, anchor=west}
        ]
        \addplot[fill=colorCatA, draw=white] table[x=catA, y=tool_id] from \datatable;
        \addlegendentry{Category A: Governance}
        
        \addplot[fill=colorCatB, draw=white] table[x=catB, y=tool_id] from \datatable;
        \addlegendentry{Category B: Technical}
        
        \addplot[fill=colorCatC, draw=white, nodes near coords, point meta=explicit] table[x=catC, y=tool_id, meta=total_score] from \datatable;
        \addlegendentry{Category C: Control}
        \end{axis}
    \end{tikzpicture}
    \caption{Final privacy scores and per-category performance of AI coding assistants. The total length of each bar corresponds to the final weighted score, which is labeled on the right. The colored segments illustrate how each privacy category contributed to that total.}
    \label{fig:final-results}
\end{figure*}
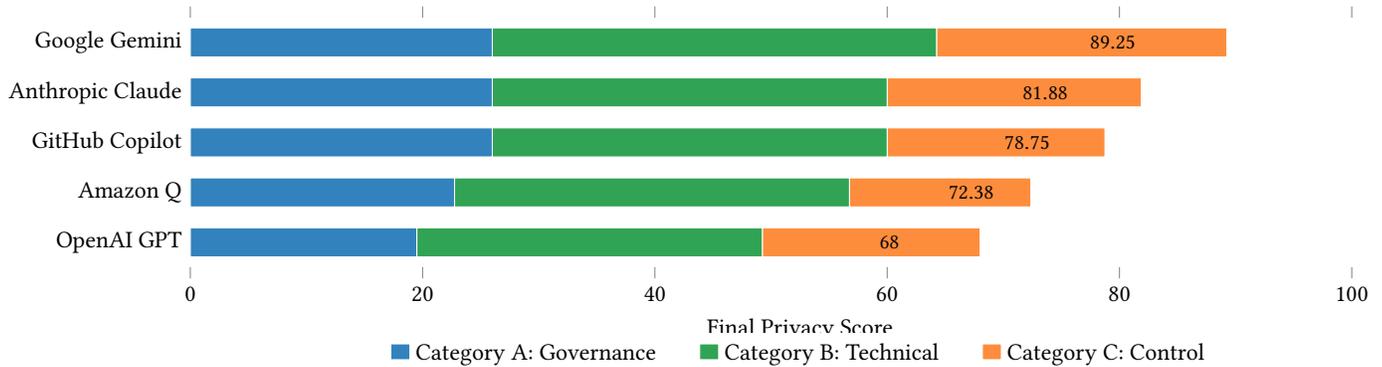

The application of the expert-validated privacy scorecard to the five selected AI coding assistants reveals significant divergence in their privacy postures. The findings highlight a landscape where stated policies can conflict with technical realities and user-protective features are inconsistently implemented. This section presents the comparative ranking, analyzes performance across key privacy dimensions, and discusses the implications for practitioners and the industry.

\subsection{Overall Privacy Rankings}
The quantitative analysis yielded a clear hierarchy of privacy performance, as visualized in Figure~\ref{fig:final-results}. \textbf{Google Gemini} emerges as the leader with a score of \textbf{89.25}, demonstrating strong performance in all categories. \textbf{Anthropic Claude (81.88)} and \textbf{GitHub Copilot (78.75)} form a competitive second tier. A significant gap separates them from \textbf{Amazon Q Developer (72.38)} and \textbf{OpenAI's GPT models (68)}, which rank last. This 20-point spread underscores that the choice of an AI coding assistant has substantial and measurable privacy implications.

\subsection{Analysis of Privacy Dimensions}
The overall scores are a composite of distinct strengths and weaknesses within each privacy category, which directly addresses the research questions.

\textbf{Data Governance \& Legal Compliance (RQ1, RQ3):} Most providers scored relatively well in this area, indicating mature baseline legal practices. GitHub Copilot led this category with near-perfect policies on paper. However, a critical weakness emerged across all providers in how they obtain consent for model training (A2). \textbf{Anthropic Claude is the sole exception, implementing a user-centric opt-in model}. All other evaluated services employ an **opt-out** model for their widely used consumer or free tiers. This default practice means user code is collected and used to improve the service, shifting the burden of privacy protection onto the user. Furthermore, the analysis of external evidence revealed that for OpenAI, legal realities (such as court-ordered data retention) contradict user-facing promises of data deletion, making their Data Subject Access Request (DSAR) mechanisms (A5) functionally misleading.

\textbf{Technical Privacy \& Security Safeguards (RQ1, RQ2, RQ3):} This category, most heavily weighted by the experts, revealed a critical cross-vendor weakness. While providers uniformly claim strong encryption (B3) and (for enterprise tiers) provide third-party audits (B5), they almost universally fail on a key developer-centric issue: proactive filtering of secrets from prompts (B4). Four out of the five assistants place the responsibility on the user to avoid inputting sensitive data like API keys or personally identifiable information (PII). As one policy warns, "Please don’t enter confidential information... or any data you wouldn’t want a reviewer to see." This answers RQ1 by showing that sensitive data, if entered, is collected and stored without proactive platform-level redaction.

\textbf{Transparency \& User Control (RQ2, RQ3):} This dimension revealed the clearest maturity gap between the leading and lower-ranked tools. Google Gemini and Anthropic Claude achieved high scores here by providing users with a suite of controls, detailed telemetry data, clear retention policies, and public model cards (C1-C4). In contrast, the other assistants lag, typically offering only a global, all-or-nothing account-level opt-out for model training and providing vague, non-numerical data retention policies (C3), stating data is kept "for as long as is necessary." This lack of specific, user-configurable controls limits a developer's ability to manage their privacy posture. This finding addresses RQ2 by highlighting the inconsistent and often inadequate nature of the privacy controls offered to users.

\subsection{Implications and Recommendations (RQ4)}
The findings of this study have direct and actionable implications for software practitioners, organizations, and tool providers.

\textbf{For Practitioners and Organizations:} The selection of an AI coding assistant must be treated as a security and compliance decision, not merely a productivity choice. The following actions are recommended:
\begin{enumerate}
    \item \textbf{Prioritize Enterprise Tiers:} Privacy protections are not uniform across service tiers. Enterprise-level subscriptions consistently offer superior safeguards. For example, enterprise agreements often include zero-data-retention policies for prompts and outputs and contractual guarantees that user data will not be used for model training. They also typically provide access to independently verified security certifications (e.g., SOC 2 reports), which are often unavailable for consumer tiers.
    \item \textbf{Assume Zero-Filtering:} Developers must operate under the assumption that any code or data pasted into an assistant may be stored and reviewed. Organizations must enforce strict policies prohibiting the inclusion of proprietary code, secrets, or PII in prompts.
    \item \textbf{Favor Opt-In by Default:} For organizations where data privacy is paramount, tools with explicit opt-in consent models should be selected. The results show Anthropic Claude is the only tool that guarantees this for all users, representing the highest standard of consent.
\end{enumerate}

\textbf{For Providers and Policymakers:} The industry has a clear path for improvement.
\begin{enumerate}
    \item \textbf{Make Opt-In the Standard:} Defaulting to using customer code for model training is a dark pattern that exploits user inertia. Opt-in consent should be the non-negotiable standard for all tiers of service.
    \item \textbf{Build Proactive Safeguards:} The burden of filtering secrets should not fall on the user. Providers must invest in reliable, automated systems to detect and redact sensitive data from prompts before processing or storage.
    \item \textbf{Embrace Radical Transparency:} Vague retention policies and missing model cards are unacceptable. All providers should follow the lead of Google and Anthropic by providing clear, numerical data retention timelines, comprehensive model cards, and user-friendly privacy dashboards.
\end{enumerate}

\section{Threats to Validity}
\label{sec:threats}

The paper acknowledges several potential threats to the validity of this study.

\textbf{Construct Validity} refers to the extent to which the scorecard accurately measures the concept of "privacy." The threat was mitigated by grounding the criteria in established privacy principles and, most importantly, by engaging a legal expert and a DPO to validate the relevance, clarity, and weighting of all criteria. This ensures the framework reflects a professional, multi-faceted understanding of privacy risk.

\textbf{Internal Validity} concerns the rigor of the scoring process. The primary threat is researcher bias in interpreting ambiguous policy language. The study addressed this by creating a detailed, evidence-based rubric with specific scoring guidelines for each sub-criterion. Furthermore, every score is linked directly to passages in the source documents, which are provided in the public replication package to allow for independent verification.

\textbf{External Validity} relates to the generalizability of the findings. The paper identifies two main limitations. First, the AI and privacy landscape is dynamic; the analysis is a \textbf{snapshot based on documents from mid-2025}. Policies, features, and even laws can change, requiring periodic re-evaluation of the tools. Second, the study is \textbf{limited to five major proprietary assistants}. The findings may not generalize to open-source, self-hosted, or niche coding assistants, which operate under different models.

\section{Conclusion}
The rapid adoption of AI coding assistants has introduced significant privacy risks due to unclear and inconsistent data handling practices. This paper addressed this challenge by developing and applying a novel, expert-validated privacy scorecard to evaluate and compare the privacy postures of five leading market providers.

The findings reveal a distinct hierarchy of privacy protections, with a 20-point gap separating the top and bottom-ranked tools. More critically, the analysis uncovered common industry weaknesses: the pervasive use of opt-out consent for model training, a near-universal failure to provide automated secret filtering, and, for some providers, a disconnect between stated policies and externally verified practices. While strong encryption is now standard, genuine user control and proactive safeguards remain the exception, not the rule.

The central contribution of this work is the scorecard framework itself—a reusable, transparent tool for holding AI service providers accountable. By translating complex legal and technical policies into a clear, quantitative measure, the framework empowers developers and organizations to make evidence-based decisions that align with their risk tolerance. Ultimately, this research serves as a call to action for greater transparency and user-centric design in AI development. The answer to "Can you trust your Copilot?" depends critically on the tool chosen, but achieving universal trust will require the entire industry to adopt the best practices highlighted in this work.

\section*{Acknowledgments}
The author wishes to express sincere gratitude to the two expert validators: \textbf{Prof. Dr. Abdulatif Alabdulatif}, a Data Privacy Officer at Qassim University, and \textbf{Rashid Alharbi}, a certified Data Architect at Saudi Health Council. Their insightful feedback and professional judgment were instrumental in shaping and validating the privacy scorecard framework at the heart of this research.

\section*{Use of AI Disclosure}
The author acknowledges the use of OpenAI's GPT-4o large language model as a writing assistant during the preparation of this manuscript. The AI's contributions included refining prose for clarity and conciseness, structuring and formatting content into the required LaTeX template, and generating the LaTeX code for the figure (Figure~\ref{fig:final-results}). However, the AI was not involved in the primary data collection, scoring, or analysis. The author conducted all original research, made all final analytical judgments, and bears full responsibility for the content and findings of this paper.

\section*{Data Availability}
The replication package for this study, is openly available on Google Sheets at: \url{https://docs.google.com/spreadsheets/d/1mpVCIv_1WSV33ovtnb-qmf1s3pRjFTonCluhGu56HDo/edit?usp=sharing}.

\bibliographystyle{ACM-Reference-Format}
\bibliography{library}

\balance 
\end{document}